\documentclass[superscriptaddress,showpacs,aps,pre,twocolumn,amsmath,amssymb,epsfig]{revtex4-1}

\usepackage{graphicx}
\usepackage{epsfig}
\usepackage{dcolumn}

\usepackage{bm}

\begin{document}

\title{Experimental Evidence for Transverse Wobbling in $^{105}$Pd}

\author {J.~Tim\'ar}
\email{Corresponding author: timar@atomki.hu}
\affiliation{Institute for Nuclear Research, Hungarian Academy of Sciences, Pf. 51, 4001 Debrecen, Hungary}

\author{Q.~B.~Chen}
\affiliation{Physik-Department, Technische Universit\"{a}t M\"{u}nchen, D-85747 Garching, Germany}

\author{B.~Kruzsicz}
\affiliation{Institute for Nuclear Research, Hungarian Academy of Sciences, Pf. 51, 4001 Debrecen, Hungary}

\author{D.~Sohler}
\affiliation{Institute for Nuclear Research, Hungarian Academy of Sciences, Pf. 51, 4001 Debrecen, Hungary}

\author{I.~Kuti}
\affiliation{Institute for Nuclear Research, Hungarian Academy of Sciences, Pf. 51, 4001 Debrecen, Hungary}

\author{S.~Q.~Zhang}
\affiliation{State Key Laboratory of Physics and Technology, School of Physics, Peking University, Beijing 100871, China}

\author{J.~Meng}
\affiliation{State Key Laboratory of Physics and Technology, School of Physics, Peking University, Beijing 100871, China}

\author{P.~Joshi}
\affiliation{Department of Physics, University of York, York, YO10 5DD, United Kingdom}

\author{R.~Wadsworth}
\affiliation{Department of Physics, University of York, York, YO10 5DD, United Kingdom}

\author{K.~Starosta}
\affiliation{Department of Chemistry, Simon Fraser University, Burnaby, British Columbia V5A 1S6, Canada}

\author{A.~Algora}
\affiliation{Institute for Nuclear Research, Hungarian Academy of Sciences, Pf. 51, 4001 Debrecen, Hungary}
\affiliation{Instituto de Fisica Corpuscular, CSIC-University of Valencia, E-46071, Valencia, Spain}

\author{P.~Bednarczyk}
\affiliation{Institute of Nuclear Physics Polish Academy of Sciences, PL-31342 Krakow, Poland}

\author{D.~Curien}
\affiliation{Universit\'e de Strasbourg, CNRS, IPHC UMR7178, 67037 Strasbourg, France}

\author{Zs.~Dombr\'adi}
\affiliation{Institute for Nuclear Research, Hungarian Academy of Sciences, Pf. 51, 4001 Debrecen, Hungary}

\author{G.~Duch\^ene}
\affiliation{Universit\'e de Strasbourg, CNRS, IPHC UMR7178, 67037 Strasbourg, France}

\author{A.~Gizon}
\affiliation{LPSC, IN2P3-CNRS/UJF, F-38026 Grenoble-Cedex, France}

\author{J.~Gizon}
\affiliation{LPSC, IN2P3-CNRS/UJF, F-38026 Grenoble-Cedex, France}

\author{D.~G.~Jenkins}
\affiliation{Department of Physics, University of York, York, YO10 5DD, United Kingdom}

\author{T.~Koike}
\affiliation{Graduate School of Science, Tohoku University, Sendai, 980-8578, Japan}

\author{A.~Krasznahorkay}
\affiliation{Institute for Nuclear Research, Hungarian Academy of Sciences, Pf. 51, 4001 Debrecen, Hungary}

\author{J.~Moln\'ar}
\affiliation{Institute for Nuclear Research, Hungarian Academy of Sciences, Pf. 51, 4001 Debrecen, Hungary}

\author{B.~M.~Nyak\'o}
\affiliation{Institute for Nuclear Research, Hungarian Academy of Sciences, Pf. 51, 4001 Debrecen, Hungary}

\author{E.~S.~Paul}
\affiliation{Oliver Lodge Laboratory, Department of Physics, University of Liverpool, Liverpool L69 7ZE, United Kingdom}

\author{G.~Rainovski}
\affiliation{Faculty of Physics, St. Kliment Ohridski University of Sofia, 1164 Sofia, Bulgaria}

\author{J.~N.~Scheurer}
\affiliation{Universit\'e Bordeaux 1, IN2P3- CENBG - Le Haut-Vigneau BP120 33175, Gradignan Cedex, France}

\author{A.~J.~Simons}
\affiliation{Department of Physics, University of York, York, YO10 5DD, United Kingdom}

\author{C.~Vaman}
\affiliation{Department of Physics and Astronomy, SUNY, Stony Brook,~New York,~11794-3800, USA}

\author{L.~Zolnai}
\affiliation{Institute for Nuclear Research, Hungarian Academy of Sciences, Pf. 51, 4001 Debrecen, Hungary}

\begin{abstract}
New rotational bands built on the $\nu$$(h_{11/2})$ configuration have been identified in $^{105}$Pd.
Two bands built on this configuration show the characteristics of transverse wobbling: the
$\Delta$$I$=1 transitions between them have a predominant E2 component and the wobbling energy decreases
with increasing spin. The properties of the observed wobbling bands are in good agreement with theoretical
results obtained using constrained triaxial covariant density functional theory and quantum particle rotor
model calculations.
This provides the first experimental evidence for transverse wobbling bands based on a one-neutron configuration,
and also represents the first observation of wobbling motion in the $A$$\sim$100 mass region.

\end{abstract}

\date{\today}
\maketitle

Nuclear wobbling motion was initially discussed by Bohr and Mottelson~\cite{BM}. This type of rotation
is predicted to occur in triaxially deformed nuclei. The nucleus rotates around the principal axis having the
largest moment of inertia and this axis executes harmonic oscillations about the space-fixed angular
momentum vector. Its analog in classical mechanics is the motion of a free asymmetric top, while in
quantal systems a corresponding example would be the rotation of molecules having different moments of 
inertia for the three principal
axes. In nuclei, the expected energy spectra related to this motion are characterized by a series of rotational
E2 bands corresponding to the different oscillation quanta (n). The signature quantum number of two
consecutive bands is different, thus the yrast and yrare bands (corresponding to n=0 and n=1, respectively),
look like signature partner bands with large signature splitting. The yrare band decays by $\Delta$$I$=1
M1+E2 transitions to the yrast band. However, contrary to the case of signature partners, the multipole mixing
ratios are very large, and the transitions have predominantly E2 character. Furthermore, the energy separation
between the yrare and yrast bands, the wobbling energy, is expected to increase with increasing spin.
Although Bohr and Mottelson predicted this motion for even-even nuclei where no intrinsic angular momentum
is involved, the phenomenon in this simple form has not been experimentally documented to date.

The first experimental evidence for nuclear wobbling motion was reported in the odd-proton $^{163}$Lu ($Z$=71)
nucleus~\cite{163Lu,163Lu2} and later in the $^{161}$Lu, $^{165}$Lu, $^{167}$Lu nuclei~\cite{161Lu,165Lu,167Lu},
as well as in $^{167}$Ta ($Z$=73)~\cite{167Ta}.
In these nuclei the wobbling mode is observed in the triaxial strongly deformed bands corresponding to
the $\pi(i_{13/2})$ intruder configuration. Recently, wobbling motion was reported in $^{135}$Pr ($Z$=59), where
the wobbling bands have normal deformation and they are built on the $\pi(h_{11/2})$ configuration~\cite{Matta2015PRL}.
The expected different signature values and the predominant E2 character of the $\Delta$I=1 transitions
between the bands have been observed for all the above cases. However, the wobbling energy has been found
to decrease with increasing spin contrary to theoretical expectations.
Frauendorf and D\"onau~\cite{Frauendorf2014PRC} interpreted this behavior as the consequence of the perpendicular
orientation of the odd particle's angular momentum to the rotational axis, and they suggested to name the
phenomenon as ``transverse wobbling". This interpretation differs from that previously published for the
Lu and Ta isotopes, and generated great theoretical interest to clarify the situation using different
models~\cite{QBC,QBC2,QBC3,TS,Raduta,Frauendorf2,TS2,Shimada,Budaca,Streck2018arXiv}. Very recently another type
of the wobbling motion has been claimed in $^{133}$La, the ``longitudinal wobbling", where the wobbling energy was 
found to increase with increasing spin \cite{Biswas}.
It is worth noting that all the wobbling bands observed so far correspond
to a configuration of one proton coupled to the core. 
In this Letter, we report on experimental evidence for transverse wobbling motion in $^{105}$Pd ($Z$=46, $N$=59).
This is the first observation of transverse wobbling motion based on a one-neutron configuration, and also the first
observation of wobbling motion in the $A$$\sim$100 mass region.

High-spin states in $^{105}$Pd were populated using the $^{96}$Zr($^{13}$C,4n) reaction. The $^{13}$C beam
was provided by the Vivitron accelerator at IReS, Strasbourg. The beam impinged upon a stack of two self
supporting metallic foil targets being enriched to 86$\%$ in $^{96}$Zr, and each having a thickness of
$\sim$0.6 mg/cm$^2$. The emitted $\gamma$-rays were detected by the EUROBALL IV~\cite{EB4} spectrometer equipped with
15 Cluster detectors at backward angles and 24 Clover detectors at 90$^{\circ}$ relative to the beam direction.
Contaminants from the charged-particle reaction channels were eliminated using the highly efficient DIAMANT
charged-particle detector array consisting of 88 CsI detectors~\cite{jns,gal} as an off-line veto. A total of
$\sim$2~$\times$~10$^9$ triple- and higher-fold coincidence events were obtained and stored onto magnetic tapes.

\begin{figure}
\begin{center}
\includegraphics[angle=-90,width=10.0cm,bb=30 160 570 750]{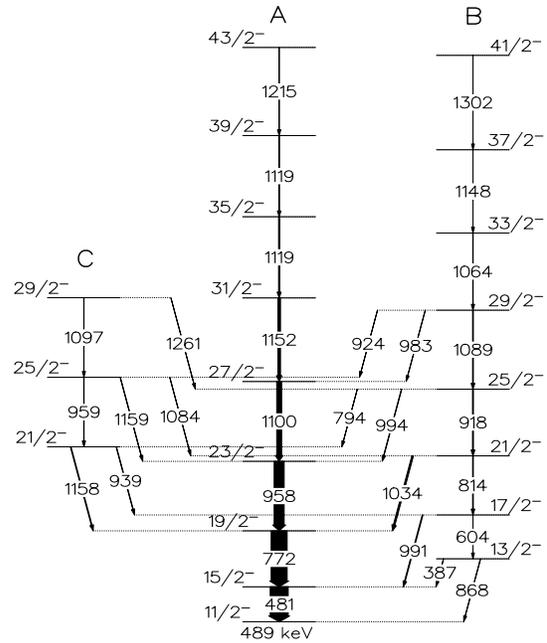}
\caption{Partial level scheme of $^{105}$Pd observed in the present work and relevant to the focus
of the present Letter. Widths of the lines are proportional with the transition intensities.}
\label{lev}
\end{center}
\end{figure}

The level scheme of $^{105}$Pd was constructed using the Radware analysis package~\cite{rad} on the
basis of the triple-coincidence relations, as well as energy and intensity balances of the observed
$\gamma$ rays. Several new rotational bands have been observed in $^{105}$Pd. Among them there are
negative-parity quadrupole bands with probable neutron $h$$_{11/2}$ configuration. Two of these
bands have opposite signature than the previously known, yrast neutron $h$$_{11/2}$ band.
Fig.~\ref{lev} shows the yrast neutron $h$$_{11/2}$ band (band A) up to spin 43/2~$\hbar$ and the two newly
identified bands (bands B and C).

\begin{figure}
\begin{center}
\bigskip \bigskip
\includegraphics[angle=-90,width=8.0cm,bb=120 20 570 560]{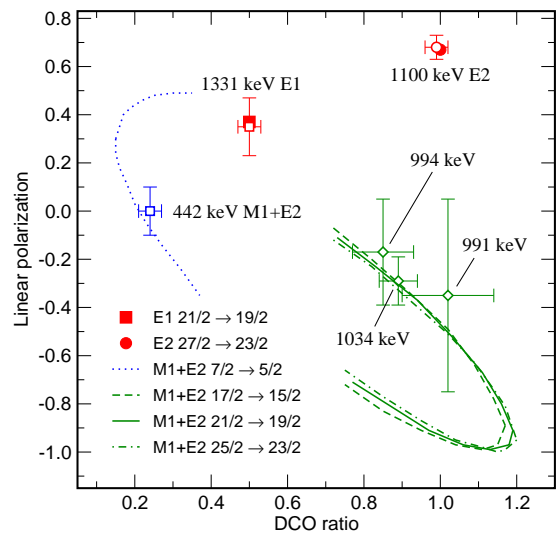}
\caption{Experimental (symbols with X and Y error bars) and calculated (full square, circle, as
well as full, dashed, dot-dashed and dotted lines) DCO and linear polarization values for the linking transitions
between bands A and B, and for three known-multipolarity transitions. In the full, dashed and dot-dashed lines
the $\delta$ multipole mixing ratio value
varies from 0.24 (lower end) to 3.2 (upper end). In the dotted line $\delta$ varies from -0.12
(lower end) to -7.1 (upper end).}
\label{mult}
\end{center}
\end{figure}

Linear polarizations and directional correlation from oriented states (DCO) ratios
\cite{pol,polD,pol1,dco} were derived for the transitions of sufficient intensity. The observed values for
the transitions relevant to the focus of the present Letter are compared in Fig.~\ref{mult} with the values
of different multipolarities and mixing ratios calculated for the experimental geometry. For the DCO ratio 
anlysis we used stretched E2 gating transitions, the attenuation
coefficients of incomplete alignment were fitted to the known strong 1100 keV E2 and 1331 keV E1
transitions~\cite{105Pd-1} in $^{105}$Pd assuming pure stretched E2 and E1 multipolarities for them, respectively.
Our analysis resulted a mixing ratio of -0.37(8) for the 442 keV lowest inband M1+E2 transition
in $^{105}$Pd which reproduced well the -0.33(13) value reported in Ref.~\cite{105Pd-1}. The 1331 keV
and 442 keV transitions are not shown in Fig.~\ref{lev}. Details of the experimental setup and data analysis, 
as well as the full level scheme, will be provided in a forthcoming publication \cite{KB}.

Band A has been reported in Refs.~\cite{105Pd-1,105Pd-2} with spin-parity values firmly assigned to
the states up to spin 31/2~$\hbar$. Data from the present experiment confirm the previously reported values.
The 17/2$^-$, 21/2$^-$ and 25/2$^-$ states of band B and the 21/2$^-$ state
of band C were reported in Ref.~\cite{105Pd-1} as non-band levels. However, the levels belonging to bands B
and C have been identified as rotational bands first in the present experiment. The DCO and linear polarization
values derived for the 814~keV, 918~keV, 1089~keV and 1064~keV transitions agree well with stretched E2 multipolarity, 
confirming the E2-band character of band B. 
As it is seen in Fig.~\ref{mult}, the measured DCO and linear polarization values for the 991 keV, 1034 keV 
and 994 keV transitions are in good agreement with $\Delta$$I$=1 M1+E2 multipolarity at large, 
$\delta$=1.8(5), 2.3(3), 2.7(6) multipole mixing ratios, respectively. Thus, these transitions have 
predominantly E2 characters, however, they cannot be pure $\Delta$$I$=2 E2 transitions because for such
transitions the linear polarization values are expected to be between 0.65 and 0.7, like in the case of the 
1100~keV gamma transition, contrarily to the measured negative values. Therefore, the observed DCO and 
polarization values allow only the 17/2$^-$, 21/2$^-$ and 25/2$^-$ spin-parity values 
for the initial states of the 991 keV, 1034 keV and 994 keV transitions, respectively. Strictly speaking, spins
less by one or two units would also be allowed by the DCO and polarization data. However, it is very rare 
that levels of rotational bands decay to the same-spin or higher-spin states of another band, and in those
cases they also decay to lower-spin levels of the other band. Existence of such transitions to lower spin states
was excluded by the observed data in the present case. 

The lowest-energy state of band C is fed by the 794 keV transition from the 25/2$^-$ state of band B
and decays by the 939 keV transition to the 17/2$^-$ state of band B. As the M2, M3 and E3 transitions are 
not competitive with E2 and M1 transitions, the 794 keV and the 939 keV transitions can only be stretched E2 
transitions. Thus the spin-parity of the lowest-energy state of band C can only be 21/2$^-$. Similarly, the 
second state of band C is linked to the 21/2$^-$ and to the 29/2$^-$ states of band B. Thus, its spin-parity 
can only be 25/2$^-$. This also confirms the E2-band character of band C. The adopted spin-parity assignments 
for the four previously known levels of bands B and C are consistent with those reported in Ref.~\cite{105Pd-1}.

Band C decays to band A by the 1158 keV and 1159 keV transitions. Linear polarization value of -0.6(3) was
derived for the sum of the two transitions. Unfortunately, no linear polarization values could be 
deduced separately for the two transitions because of their close energies. DCO value could not be derived 
even for the sum of the two transitions because their energies are close to that of the strong 1152 keV transition 
in band A. The fact that the 1158 keV, 1159 keV and 1152 keV transitions
are all in coincidence with the intense gamma rays which could be used as coincidence gates, caused further
difficulties in the analysis. The deduced linear polarization value agrees well with the multipolarity expected
for the 1158 keV and 1159 keV transitions from the above spin-parity assignments for the band C states: namely
that they are $\Delta$$I$=1 M1+E2 transitions. However, it allows both small (0 $\le$ $\delta$ $\le$ 0.5) and
large (1 $\le$ $\delta$ $\le$ 2.4) mixing ratios.

The observed three bands show the features of a pair of wobbling bands with oscillation
quanta zero and one (bands A and B, respectively) and the signature partner band of band A (band C).
Indeed, the multipolarities of the lowest-lying linking transitions between bands B and A are M1+E2
with large, $\delta$=1.8(5), 2.3(3), 2.7(6) multipole mixing ratios for the 991 keV, 1034 keV and 994 keV 
transitions, respectively. These mixing ratios mean around 80$\%$ (calculated as $\delta^2/(1+\delta^2)$) E2
content, which is expected for the wobbling band, but not expected for the signature partner. We note
that the 991 keV, 1034 keV and 994 keV transitions were also reported in Ref.~\cite{105Pd-1} and $\delta$=0.46(10)
as well as $\delta$=0.62(18) were derived for the 991 keV and 1034 keV transitions, respectively, from
angular distribution measurement. While the present DCO results also allow $\delta$=0.59(20) and 0.40(6) 
values for the two transitions, respectively, the linear polarization data disagree with these smaller 
mixing ratios, but strongly support the larger $\delta$=1.8(5) and 2.3(3) values.

Band C is a candidate for the signature partner of band~A. The two bands have the same parity and similar
alignments~\cite{KB} but opposite signature. Furthermore, band C decays to band A by the 1158 keV and 
1159 keV transitions. 
Although the mixing ratios of these transitions could not be derived unambiguously, the possible smaller
mixing ratio value deduced from the present experiment is in a good agreement with this scenario.
In Ref.~\cite{105Pd-1} a mixing ratio of $\delta$=1.3(9) was reported for the 1158 keV transition.
Due to the large uncertainty this value can allow a rather small mixing ratio, thus it can be in
agreement with the signature partner interpretation, too. 

The difference between the mixing ratio values measured for the linking M1+E2 transitions between the
wobbling bands in $^{135}$Pr and in $^{105}$Pd is their opposite signs. While the sign is positive
in $^{105}$Pd, it is negative in $^{135}$Pr. The sign of the mixing ratio value is determined by the
sign of the M1 matrix element assuming that the quadrupole deformation is of same type.
The sign of the M1 matrix element is proportional with the ($g_p-g_R$) factor~\cite{BM}, where $g_R$
is the rotational gyromagnetic factor. Its value is approximately $Z/A$, which is $\sim$0.4 for both nuclei.
However, $g_p$, the gyromagnetic factor of the odd particle, is different for the protons and the
neutrons moving in high-$j$ intruder ($j=l+1/2$) orbitals. It has a large positive value ($>$1) for protons, while it has a
negative sign for neutrons. Thus, the sign of the ($g_p-g_R$) factor is opposite for high-$j$ protons and
neutrons \cite{Nakai}.

In order to explore the nature of the observed rotational band structures in $^{105}$Pd, they have been studied
by the constrained triaxial covariant density functional theory
(CDFT)~\cite{J.Meng2006PRC, J.Meng2016book} as well as the quantum particle
rotor model (PRM)~\cite{FM1, FM2, Hamamoto2002PRC, Frauendorf2014PRC, W.X.Shi2015CPC,
Streck2018arXiv}.
The configuration-fixed CDFT calculations~\cite{J.Meng2006PRC, J.Meng2016book} with the effective 
interaction PC-PK1~\cite{P.W.Zhao2010PRC} reveal that the
$\nu(1h_{11/2})^1$ configuration has a triaxial shape of $\beta=0.27$
and $\gamma=25^\circ$, which fulfills the conditions required for the presence of wobbling bands.

\begin{figure}[!ht]
  \begin{center}
    \includegraphics[width=8.0 cm]{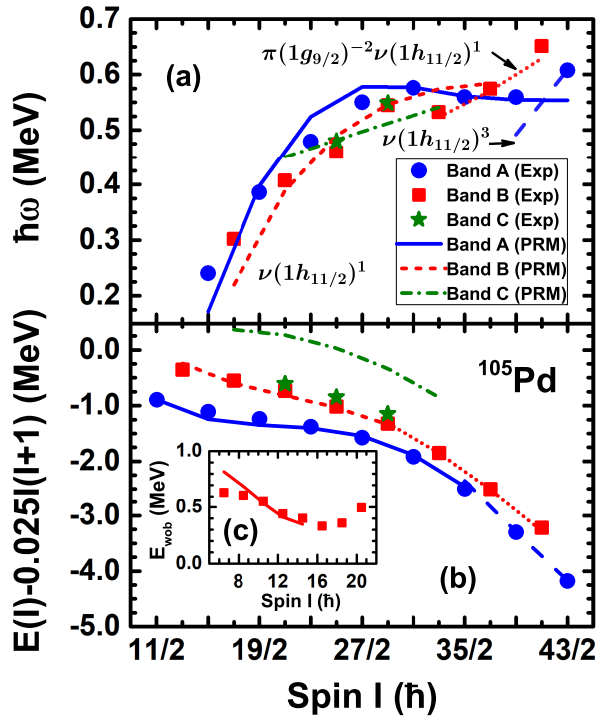}
    \caption{Experimental and PRM rotational frequency (a) as well as energies minus
    a rotor contribution (b) as functions of spin $I$ for the bands A, B, and C in
    $^{105}$Pd. Inset (c): Wobbling energies associated with the wobbler-band pair
    A and B.}\label{fig4}
  \end{center}
\end{figure}

With the configuration and deformation parameters obtained, it is straightforward 
to perform PRM~\cite{Hamamoto2002PRC,
Frauendorf2014PRC, W.X.Shi2015CPC, Streck2018arXiv} calculations in order to study the energy
spectra and electromagnetic transition probabilities for the observed
rotational sequences in $^{105}\rm Pd$. In the PRM calculations, the neutron particle
is described by a single-$j$ shell Hamiltonian~\cite{Ring1980book} and the
pairing effect is included using the standard BCS quasiparticle approximation
with the empirical pairing gap $\Delta=12/\sqrt{A}=1.17~\textrm{MeV}$ and the
Fermi surface located at the beginning of the $h_{11/2}$ subshell.
The triaxial rotor is parametrized by three angular-momentum-dependent moments of inertia
$\mathcal{J}_i=a_i\sqrt{1+bI(I+1)}$~\cite{C.S.Wu1985HEPNP, C.S.wu1987CTP}
(known as the $ab$ formula) to take into account the soft character of
the potential energy surface revealed by CDFT calculations~\cite{Y.Y.Wang2016PRC}. 
Here, $i=m$, $s$, $l$ denotes the medium,
short, and long axes, respectively, and the corresponding parameters
$a_{m,s,l}=5.89, 3.74, 1.27~\hbar^2/\textrm{MeV}$ and $b=0.023~\hbar^{-2}$ are
determined by fitting to the energy spectra of bands A~and~B.

The calculated rotational frequency ($\hbar\omega(I)=[E(I)-E(I-2)]/2$) and energy spectra as functions of
spin $I$ for bands A (solid line), B (short dash line), and C (short dash-dot
line), in comparison with those of the experimental data, are shown
in Fig.~\ref{fig4}. It is seen that the PRM calculations can
reproduce bands A and B well. For band C, the energies are
overestimated by about 500 keV. A similar problem is also seen in
Ref.~\cite{Matta2015PRL} for $^{135}$Pr.

\begin{table*}[!ht]
\caption{The experimental and theoretical multipole mixing ratios $\delta$ as well
as the transition probability ratios $B(M1)_{\textrm{out}}/B(E2)_{\textrm{in}}$
and $B(E2)_{\textrm{out}}/B(E2)_{\textrm{in}}$ for the
transitions from band B to A in $^{105}$Pd.}
\label{tab1}
\begin{tabular}{cccccccccccc}
\hline
\hline
& & \multicolumn{2}{c}{$\delta$}
& \multicolumn{2}{c}{$\frac{B(M1)_{\textrm{out}}}{B(E2)_{\textrm{in}}}(\frac{\mu_N^2}{e^2b^2})$}
& \multicolumn{2}{c}{$\frac{B(E2)_{\textrm{out}}}{B(E2)_{\textrm{in}}}$} \\
$I_i^\pi~~ \to ~~I_f^\pi$ & $E_\gamma$ (keV) & Exp  & ~~PRM~~ & Exp  & ~~PRM~~ & Exp  & ~~PRM~~ \\
\hline
$17/2^- \to 15/2^-$ & 991 & $1.8\pm 0.5$  & 2.38 & $0.162\pm 0.097$ & 0.105 & $0.66\pm 0.18$ & 0.736\\
$21/2^- \to 19/2^-$ & 1034 & $2.3\pm 0.3$ & 2.30$\footnote{Normalization point, see text.}$ & $0.089\pm 0.026$ & 0.069 & $0.60\pm 0.09$ & 0.465\\
$25/2^- \to 23/2^-$ & 994 & $2.7\pm 0.6$  & 1.99 & $0.029\pm 0.016$ & 0.057 & $0.34\pm 0.07$ & 0.329\\
\hline
\end{tabular}
\end{table*}

For band A, the rotational frequency is almost constant from spin
$I=27/2$ to $39/2$, which presents an upbend phenomenon and is understood
by the gradual alignment of a $h_{11/2}$ neutron pair. Such an
alignment process can be reproduced by the PRM calculations due to the use of
angular-momentum-dependent moments of inertia. After the upbending,
the configuration becomes a three-quasiparticle configuration $\nu(1h_{11/2})^3$,
whose quadrupole deformation parameters are $\beta=0.29$ and $\gamma=10^\circ$
from the CDFT calculations, and the data can be reproduced by the PRM (dash line) with the
moments of inertia taken as irrotational flow type $\mathcal{J}_k=\mathcal{J}_0
\sin^2(\gamma-2k\pi/3)$ with $\mathcal{J}_0=20~\hbar^2/\textrm{MeV}$.

For band B, the experimental rotational frequency has a discontinuity between $I=29/2$ and
$33/2$, which is understood by the alignment of a proton $g_{9/2}$ pair given
that its alignment is $2\hbar$ smaller than that of band A in the region
$I\geq 39/2$. Hence, the unpaired valence nucleon configuration for band B
at $I\geq 33/2$ is assigned as $\pi(1g_{9/2})^{-2}\otimes\nu(1h_{11/2})^1$,
whose deformation parameters are $\beta=0.25$ and $\gamma=28^\circ$ according
to the CDFT calculations. With this configuration and $\mathcal{J}_0=21~\hbar^2/\textrm{MeV}$,
the corresponding experimental rotational frequencies and energies can
be well reproduced as shown in Fig.~\ref{fig4} (short dotted line), and thus
supports the configuration assignment.

With the successful reproduction of the energy spectra of bands A and B, the
wobbling energy $E_{\textrm{wob}}$ (as defined in Ref.~\cite{Matta2015PRL}) can also be 
reproduced by the PRM calculations, as shown in Fig.~\ref{fig4}(c). In agreement with
the experimental observation, the calculated wobbling energy 
decreases with spin until $I=29/2$, which presents the characteristic
of a transverse wobbler. Note that the increasing energy
difference between bands A and B in the region $I\geq 33/2$
cannot be interpreted as evidence of a longitudinal wobbler~\cite{Frauendorf2014PRC}, since
their configurations are different as discussed above.

In Table~\ref{tab1}, the experimental and theoretical mixing ratios $\delta$ as well
as the transition probability ratios $B(M1)_{\textrm{out}}/B(E2)_{\textrm{in}}$
and $B(E2)_{\textrm{out}}/B(E2)_{\textrm{in}}$ for the transitions from band B
to A in $^{105}$Pd are listed.
It is known that $B(E2)_{\textrm{out}}/B(E2)_{\textrm{in}}$ is proportional to
$\tan^2\gamma$~\cite{BM, TS}. It is found that the PRM results are in good agreement with the data.
Thus, the microscopic input of the triaxial deformation parameter from the CDFT
calculation is appropriate.

The mixing ratios $\delta$ and $B(M1)_{\textrm{out}}/B(E2)_{\textrm{in}}$ are
proportional to $Q_0/g_{\textrm{eff}}$ and
$(g_{\textrm{eff}}/Q_0)^2$, respectively, with $Q_0$ the intrinsic
quadrupole moment and $g_{\textrm{eff}}=g_{\nu h_{11/2}}-g_R$ the
effective gyromagnetic factor. It was found that in the PRM
calculations, the $B(M1)_{\textrm{out}}$ values would be
overestimated by about a factor of 3$-$10~\cite{Frauendorf2014PRC,
Matta2015PRL}. This is due to the scissors mode which is mixed with the wobbling 
motion and cannot be considered in the PRM calculations~\cite{Frauendorf2015PRC}.
Bearing this in mind, a
quenching factor of 0.36 for $g_{\textrm{eff}}$ is introduced in the
calculation in order to reproduce the value of $\delta$ for the
transition $21/2^- \to 19/2^-$. With this treatment, the other
experimental $\delta$ values as well as the
$B(M1)_{\textrm{out}}/B(E2)_{\textrm{in}}$ values can also be
reproduced. The large $B(E2)_{\textrm{out}}/B(E2)_{\textrm{in}}$ and small 
$B(M1)_{\textrm{out}}/B(E2)_{\textrm{in}}$ values further
support the wobbling interpretation for the bands A and B in the
region $I\leq 29/2$.

With the successful reproduction of the energy spectra and electromagnetic transitions in $^{105}$Pd,
the angular momentum geometries of bands A and B are examined in the PRM~\cite{Streck2018arXiv}.
Indeed, it is found that the neutron angular momenta in bands A and B are similar and mainly align 
along the $s$-axis in the region $I\leq 25/2$. The wave functions of band A are symmetric and peaked 
at the $s$-axis below
$I=23/2$, which indicates a principal axis rotation around the $s$-axis. At $I=29/2$,
the rotational mode of band A changes to a planar rotation. With one-phonon excitation, the wave
functions of band B are antisymmetric and have a node around the $s$-axis. These features are the same
as in the case of $^{135}$Pr~\cite{Streck2018arXiv}, thus further confirming the transverse wobbling
interpretation for bands A and B of $^{105}$Pd in the region $I\leq 29/2$.

In summary, we have studied nuclear transverse wobbling in $^{105}$Pd, where the wobbling bands
are based on the $\nu$$(h_{11/2})$ one-neutron configuration.
The predominant E2 character of the $\Delta$$I$=1 M1+E2 transitions between the wobbling bands is
confirmed by the precise measurement of DCO values and linear polarization data. The transverse
wobbling nature of these bands conforms well to results from calculations using constrained triaxial
covariant density functional theory and the quantum particle rotor model.
This observation provides the first experimental evidence for transverse wobbling bands based on a
one-neutron configuration, and is also the first observation of wobbling motion in the $A$$\sim$100
mass region.

The authors thank Professor C. M. Petrache for stimulating and useful discussions.
This work was supported by the National Research, Development and Innovation Fund of Hungary,
financed under the K18 funding scheme with project nos. K128947 and K124810.
This work was also supported by the European Regional Development Fund (Contract No.
GINOP-2.3.3-15-2016-00034), as well as by Deutsche Forschungsgemeinschaft (DFG) and National 
Natural Science Foundation of China (NSFC) through funds provided to the Sino-German CRC 110
``Symmetries and the Emergence of Structure in QCD'', the National Key R\&D Program of China 
(Contract No. 2018YFA0404400), the NSFC under Grants No. 11335002 and No. 11621131001, the 
UK STFC under grant no. ST/P003885/1, and the Spanish Ministerio de Economia y Competitividad 
under Grant No. FPA2014-52823-C2-1-P and the program Severo Ochoa (SEV-2014-0398). 
I.K. was supported by National Research, Development and Innovation Office – NKFIH, contract 
number PD 124717.

\end{document}